\documentclass[12pt]{article}
\usepackage{latexsym,amssymb,amsmath}
\usepackage{amsthm, amstext}
\usepackage{array, amsfonts, mathrsfs}
\usepackage[dvips]{graphicx}
\usepackage[dvips]{graphicx,color}

\newtheorem{Proposition}{Proposition}

\numberwithin{equation}{section}

\date{}
\begin{document}

\author{M.I.Belishev\thanks {Saint-Petersburg State University, 7/9 Universitetskaya
                 nab., St. Petersburg, 199034 Russia, m.belishev@spbu.ru.
                 Saint-Petersburg Department of the Steklov Mathematical Institute,
                 belishev@pdmi.ras.ru. Supported by the grants RFBR 14-01-00535À and
                 Volks\-Wagen Foundation"}.}

\title{On algebras of three-dimensional quaternionic harmonic fields}
\maketitle

\begin{abstract}
A quaternionic field is a pair $p=\{\alpha,u\}$ of function
$\alpha$ and vector field $u$ given on a 3d Riemannian maifold
$\Omega$ with the boundary. The field is said to be harmonic if
$\nabla \alpha={\rm rot\,}u$\, in $\Omega$. The linear space of
harmonic fields is not an algebra w.r.t. quaternion
multiplication. However, it may contain the commutative algebras,
what is the subject of the paper. Possible application of these
algebras to the impedance tomography problem is touched on.
\end{abstract}


\bigskip

{\rightline {\bf In memory of Gennadii Markovich Henkin}}

\setcounter{section}{-1}
\section{Introduction}\label{sec Introduction}
{\bf Motivation.}\,\,\, Let $({\Omega},g)$ be a
smooth\footnote{everywhere in the paper `smooth' means
$C^\infty$-smooth} compact Riemannian manifold with the boundary
${\Gamma}$, $\Delta_g$ the Beltrami-Laplace operator, $u=u^f(x)$ a
solution of the problem
 \begin{align*}
& \Delta_g u = 0 && {\rm in}\,\,\, {\Omega}\\
& u = f  && {\rm on}\,\,\, {\Gamma},
 \end{align*}
$\Lambda: f \mapsto \partial_\nu u^f|_{\Gamma}$ the
Dirichlet-to-Neumann operator, $\nu$ the outward normal on
${\Gamma}$. An {\it impedance tomography problem} (ITP) is to
recover $({\Omega},g)$ via the given $\Lambda$.

For the case ${\rm dim\,}\Omega=2$, an algebraic approach to ITP
is proposed in \cite{BCald}. Its key device is the commutative
Banach algebra of analytic functions
 \begin{equation}\label{CR 2d}
{\cal A}({\Omega})\,:=\,\left\{w={\varphi}+\psi i\,|\,\,{\varphi},
\psi \in C({\Omega}),\,d\psi = \star \,d{\varphi} \,\,{\rm
in}\,\,\,{\Omega} \setminus {\Gamma}\right\}\,.
 \end{equation}
The Gelfand spectrum $\widehat{{\cal
A}({\Omega})}=:{\Omega}^{\mathbb C}$ (the set of homomorphisms
${\cal A}({\Omega})\to {\mathbb C}$) of this algebra is
homeomorphic to the manifold: ${\Omega}^{\mathbb C}\cong\Omega$.
The algebra of boundary values ${\cal
A}({\Gamma})=\{w|_{\Gamma}\,\,|\,\,w \in {\cal A}({\Omega})\}$ is
isometrically isomorphic to ${\cal A}({\Omega})$. Therefore, the
spectra of these algebras are canonically homeomorphic:
$\widehat{{\cal A}({\Gamma})}\cong{\Omega}^{\mathbb C}$. In the
mean time, ${\cal A}({\Gamma})$ is determined by the DN-operator
$\Lambda$. The latter enables one to recover ${\Omega}$ up to
homeomorphism by the scheme $\Lambda \Rightarrow {\cal
A}({\Gamma})\Rightarrow \widehat{{\cal
A}({\Gamma})}\cong{\Omega}^{\mathbb C}\cong{\Omega}$ (see
\cite{BCald, BIP'07} for more detail).

For ${\dim\,}{\Omega} \geqslant 3$, the known results on ITP
\cite{FKLS,KSU,LTU} concern to a certain specific class of
admissible metrics $g$. In the mean time, the attempt to extend
the algebraic approach encounters the following obstacle. The
relevant multidimensional analog of ${\cal A}({\Omega})$ is the
space of differential forms satisfying the Cauchy-Riemann
condition $d\psi = \star \,d{\varphi}$ \cite{BCUBO, BShar}.
Unfortunately, this space is not an algebra. However, at least in
the 3d-case, it may possess of certain algebraic properties, which
are the subject of our paper. We hope for utility of these
properties for the future progress in ITP.
\bigskip

\noindent{\bf Contents.}\,\,\, We deal with the case
${\dim\,}{\Omega} = 3$. A {\it quaternion field} on $\Omega$ is a
pair $p=\{\alpha,u\}$ of (real valued) function $\alpha$ and
vector field (section of $T{\Omega}$) $u$. The space ${\cal
C}({\Omega})=\{p\,|\,\,\alpha \in C(\Omega),\,u \in \vec
C(\Omega)\}$ with the ${\rm sup}$-norm is a noncommutative Banach
algebra w.r.t. the multiplication $pq=\{\alpha \beta - u\cdot
v,\alpha v + \beta u +u\wedge v\}$, where $q=\{\beta,v\}$, $\cdot$
and $\wedge$ are the point-wise inner and vector products in the
tangent spaces $T\Omega_m$. This space contains the (sub)space of
{\it harmonic fields} ${{\cal Q}(\Omega)}=\{p \in {\cal
C}({\Omega})\,|\,\,\nabla \alpha={\rm rot\,} u\}$, which is not a
(sub)algebra: generically $p,q \in {{\cal Q}(\Omega)}$ does not
imply $pq \in {{\cal Q}(\Omega)}$.

However, we show that, under certain conditions on the metric $g$,
the space ${\cal Q}(\Omega)$ may contain the commutative algebras
${\cal A}_e(\Omega)$ associated with the geodesic fields $e$ and
similar to the above mentioned 2d algebras ${\cal A}({\Omega})$.
These algebras determine a {\it quaternionic spectrum}
$\Omega^{\mathbb H}$, which is a candidate for the role of
relevant 3d analog of the 2d Gelfand spectrum ${\Omega}^{\mathbb
C}$.

From the viewpoint of Algebra, the case ${\Omega} \in {\mathbb
R}^3$ is richer in content: the space ${{\cal Q}(\Omega)}$
contains a subspace ${\dot{\cal Q}(\Omega)}$, which is an
AH-module \cite{Joyce}, whereas the spectrum ${\Omega}^{\mathbb
H}$ is well defined and homeomorphic to $\Omega$.

It is the possible homeomorphism $\Omega^{\mathbb H}\cong\Omega$,
which enables us to hope for application of the quaternionic
spectrum to the 3d ITP. We mean the reconstruction of $\Omega$ by
the scheme: $\Lambda\Rightarrow$ an isometric copy $\tilde{\cal
Q}(\Omega)$ of the space ${{\cal Q}(\Omega)} \Rightarrow$ its
spectrum ${\tilde \Omega}^{\mathbb H} \cong \Omega^{\mathbb
H}\cong\Omega$.
\bigskip

\noindent{\bf Acknowledgements.}\,\,\,Working on this paper, I
used helpful consultations of my colleagues. G.M.Henkin has
acquainted me with the article \cite{Joyce} by D.Joyce. I was able
to understand its content owing to elucidation and comments of
Yu.A.Kor\-dyu\-kov. L.N.Pestov and S.V.Ivanov consulted me in
questions on geometry. S.Vessella provided me with results and
references on elliptic PDEs. I'm extremely grateful to all of them
for the kind help. I'd like to thank M.Salo for information and
references on the papers \cite{FKLS,KSU}.

\section{Harmonic quaternion fields}
{\bf Quaternions.}\,\,\,$\bullet$\,\,\,By ${\mathbb H}
=\{q=a+P{\bf i}+Q{\bf j}+R{\bf k}\,|\,\,a,P,Q,R \in \mathbb R\}$
we denote the {\it quaternion algebra} endowed with the standard
linear operations and multiplication determined by the table ${\bf
i}^2={\bf j}^2={\bf k}^2=-1;\,\,{\bf i}{\bf j}={\bf k},\,\,{\bf
j}{\bf k}={\bf i},\,\,{\bf k}{\bf i}={\bf j}$ (see, e.g,
\cite{Korn}). Also, one defines the involution $q\mapsto\bar
q=a-P{\bf i}-Q{\bf j}-R{\bf k}$ and modulus $|q|=(q\bar
q)^{\frac{1}{2}}=\left(a^2+P^2+Q^2+R^2\right)^{\frac{1}{2}}$. We
denote $\Re q=a,\,\,\Im q=P{\bf i}+Q{\bf j}+R{\bf k}$, and call
elements of the subspace ${\mathbb I}=\{q\in {\mathbb H}\,|\,\,\Re
q=0\}$ the imaginary quaternions.

\noindent$\bullet$\,\,\, Let $E$ be a real oriented 3-dimensional
Euclidean space, $\cdot$ and $\wedge$ the inner and vector
products in $E$. A pair $q=\{\alpha,u\}$ with $\alpha\in{\mathbb
R}$ and $u \in E$ is said to be a geometric quaternion. We denote
$\Re q=\alpha,\,\,\Im q=u$.

The 4d linear space ${\cal C}$ of such pairs endowed with the
compo\-nent-wise summation, multiplication
 \begin{equation}\label{multiplication}
qp:=\{\alpha \beta - u\cdot v,\alpha v + \beta u +u\wedge v\}
 \end{equation}
(here $p=\{\beta,v\}$), involution $q\mapsto \bar
q=\{\alpha,-u\}$, and modulus $|q|=(q\bar
q)^{\frac{1}{2}}=\left(\alpha^2+|u|^2_E\right)^{\frac{1}{2}}$ is
an algebra.  Elements of the subspace ${\cal I}=\{q\in {\mathbb
H}\,|\,\,\Re q=0\}$ are named by imaginary (geometric)
quaternions.

Choosing an orthonormal basis $e_1, e_2, e_3 \in E$ and
representing $u=Ae_1+Be_2+Ce_3$, one determines the isometric
isomorphism between the algebras ${\cal C}$ and ${\mathbb H}$ by
$\{\alpha, Ae_1+Be_2+Ce_3\} \leftrightarrow \alpha+A{\bf i}+B{\bf
j}+C{\bf k}$, the isomorphism mapping ${\cal I}$ onto ${\mathbb
I}$. By this, we identify ${\cal C}$ and ${\mathbb H}$.

\noindent$\bullet$\,\,\, As well as ${\mathbb H}$, algebra ${\cal
C}$ is noncommutative. However, it contains commutative
subalgebras of the form
 \begin{align}
\notag & {\cal A}_e=\{p=\{{\varphi},\psi e\}\,|\,\,{\varphi},\psi
\in {\mathbb
R};\,\,e \in E,\,\,|e|=1\}\,;\\
\label{A-h in C} & {\cal A}_0\,=\,\{q=\{\alpha,0\}\,|\,\,\alpha
\in {\mathbb R}\}\,,
 \end{align}
which are isometrically isomorphic to ${\mathbb C}$ (by ${\cal C}
\ni p \leftrightarrow {\varphi}+\psi{\bf i}\in {\mathbb C}$) and
$\mathbb R$ respectively. As is easy to see, any commutative
subalgebra in ${\cal C}$ is of the form (\ref{A-h in C}). Indeed,
if $p,q \in {\cal C}$ and $pq=qp$ holds then $\Im p \wedge \Im
q\overset{(\ref{multiplication})}=0$, i.e., $\Im p$ and $\Im q$
has to be linearly dependent.
\bigskip

\noindent{\bf Vector analysis.}\,\,\,Let $\Omega$ be a smooth
oriented Riemannian manifold, ${\rm dim\,}{\Omega}=3$, $g$ the
metric tensor, $\mu$ the Riemannian volume 3-form, $\star$ the
Hodge operator, $\nabla_u$ the covariant derivative. On such a
manifold, the intrinsic operations of vector analysis are well
defined on smooth functions and vector fields (sections of the
tangent bundle $T\Omega$). Following \cite{Sch}, we recall their
definitions.
\smallskip

\noindent{$\bullet$}\,\,\,For a field $u$, one defines the {\it
conjugate $1$-form} $u_\sharp$ by $u_\sharp(v)=g(u,v),\,\,\forall
v$. For a $1$-form $f$, the {\it conjugate field} $f^\sharp$ is
defined by $g(f^\sharp,u)=f(u),\,\,\forall u$.

\noindent{$\bullet$}\,\,\,The {\it scalar product} $\,\cdot\,\,:\,
\{ {\rm fields}\} \times \{ {\rm fields}\} \, \to \, \{ {\rm
functions}\}$ is defined point-wise by $u\cdot v=g(u,v)$. The {\it
vector product} $\wedge :\, \{ {\rm fields}\} \times \{ {\rm
fields}\} \, \to \, \{ {\rm fields}\}$ is defined point-wise by
$g(u \wedge v,w)=\mu\,(u,v,w),\,\forall w$.

\noindent{$\bullet$}\,\,\,The {\it gradient} $\nabla :\, \{ {\rm
functions}\} \to \{ {\rm fields}\}$ and {\it divergence} ${\rm
div\,}:\, \{ {\rm fields}\} \to \{ {\rm functions}\}$ are defined
by $\nabla \alpha=(d\alpha)^\sharp$ and ${\rm div}\,u=\star
d\!\!\star u_\sharp$ respectively, where $d$ is the exterior
derivative.

\noindent{$\bullet$}\,\,\,The {\it rotor} maps $\{ {\rm fields}\}$
to $\{ {\rm fields}\}$ by ${{\rm rot\,}}u =(\star d\,u_\sharp
)^\sharp$. Recall the basic identities: ${\rm div}\,{{\rm
rot\,}}=0$ and ${\rm rot\,} \nabla =0$. The equalities
 \begin{equation*}
\nabla\alpha\,=\,{\rm rot\,} u \qquad {\rm and}\qquad
d\alpha=\star\,du_\sharp
  \end{equation*}
are equivalent. By analogy with the Cauchy-Riemann conditions in
(\ref{CR 2d}), they are called the CR-conditions.

\noindent{$\bullet$}\,\,\,The {\it Laplacian} $\Delta :\, \{ {\rm
functions}\} \to \{ {\rm functions}\}$ is $\Delta ={\rm
div\,}\nabla$.
\smallskip

For smooth functions $\alpha, \beta$ and fields $u,v$, the
following relations hold:
 \begin{align}
\notag &  \nabla \alpha
\beta=\beta\nabla\alpha+\alpha\nabla\beta;\,\,\,
\nabla\, u\!\cdot\! v = \nabla_vu+\nabla_uv+v\wedge{\rm rot\,} u +u \wedge{\rm rot\,} v;\\
\notag & {\rm rot\,} \alpha v=\nabla\alpha\wedge v+\alpha\,{\rm
rot\,} v;\,\,\, {\rm rot\,} (u\wedge v)=
\nabla_vu-\nabla_uv-({\rm div\,} u)v+({\rm div\,} v)u; \\
\label{F1} & {\rm div\,} u\wedge v=v\cdot{\rm rot\,} u -
u\cdot{\rm rot\,} v;\,\,\,{\rm div\,} \alpha v = \nabla\alpha\cdot
v+\alpha\,{\rm div\,} v
 \end{align}
(see \cite{Korn}, Chapter 16.8, and \cite{Sch}, Chapter 10).

In what follows we deal with a compact ${\Omega}$ with the
boundary $\Gamma$. By $C({\Omega})$ and $\vec C({\Omega})$ we
denote the Banach spaces of continuous functions and vector fields
endowed with the standard $\rm sup$-norms.
\bigskip

\noindent{\bf Quaternion fields.}\,\,\,{$\bullet$}\,\,\,A
quaternion {\it field} is a pair $q=\{\alpha,u\}$, where
$\alpha=\Re q$ and $u=\Im q$ are a function and vector field given
in $\Omega$. The space of pairs ${\cal
C}({\Omega})=\{q\,|\,\,\alpha \in C({\Omega}),\,\,u \in \vec
C(\Omega)\}$ with the point-wise summation and multiplication
(\ref{multiplication}), and the norm
 $$
\|q\|\,=\,\underset{x\in\Omega}{\rm
sup\,}|q(x)|\,=\,\underset{x\in\Omega}{\rm
sup\,}\left(|\alpha(x)|^2+
|u(x)|^2_{T\Omega_x}\right)^{\frac{1}{2}}
 $$
is a noncommutative Banach algebra; in particular,
$\|qp\|\leqslant\|q\|\|p\|$ does hold. The set of imaginary fields
${\cal I}({\Omega})=\{q \in {\cal C}({\Omega})\,|\,\,\Re q=0\}$ is
a subspace but not a subalgebra in ${\cal C}({\Omega})$.

\noindent{$\bullet$}\,\,\,Elements of the subspace
 $$
{{\cal Q}(\Omega)}\,=\,\left\{q \in {\cal
C}({\Omega})\,|\,\,\nabla\alpha\,=\,{\rm rot\,} u\,\,\,{\rm
in}\,\,\,{\Omega}\setminus\Gamma \right\}
 $$
are said to be {\it harmonic fields}. Also, we introduce the
subspace
 $$
{\dot{\cal Q}(\Omega)}\,=\,\left\{q \in {\cal
C}({\Omega})\,|\,\,\nabla\alpha\,=\,{\rm rot\,} u,\,\,{\rm div\,}
u=0\,\,\,{\rm in}\,\,\,{\Omega}\setminus\Gamma
\right\}\,\subset\,{{\cal Q}(\Omega)}
 $$
and call its elements {\it pure harmonic fields}.

\section{Axial algebras}

Neither ${\cal Q}(\Omega)$ nor ${\dot{\cal Q}(\Omega)}$ are the
(sub)algebras in ${\cal C}({\Omega})$ since, generically,
multiplication (\ref{multiplication}) does not preserve
harmonicity. However, as will be shown, under some conditions on
the manifold ${\Omega}$, these subspaces may contain commutative
algebras. These algebras are of the main our interest.
\bigskip

\noindent{\bf Formulas.}\,\,\, Let
$p=\{\alpha,u\},\,q=\{\beta,v\}$ be smooth quaternion fields in
$\Omega$. A field
 $
{\varepsilon}(p)\,=\,\nabla \alpha-{\rm rot\,} u
 $
is said do be a {\it harmonic residual} of $p$. By this
definition, one has ${\varepsilon}|_{{\cal Q}(\Omega)}=0$. Using
(\ref{F1}), one derives the following equalities:
 \begin{align}
\notag & {\varepsilon}(pq)\,=\\
\label{E1} &
=\beta{\varepsilon}(p)+\alpha{\varepsilon}(q)+v\wedge{\varepsilon}(p)+u\wedge{\varepsilon}(q)+({\rm
div\,}
 u)v-({\rm div\,} v)u - 2 \nabla_vu\,,\\
\notag & {\rm div\,} \Im(pq)={\rm div\,}(\alpha v + \beta u + u\wedge v)\,=\\
\label{E2} &=\, \alpha \,{\rm div\,} v+\beta\,{\rm div\,} u
+u\cdot{\varepsilon}(q)+v\cdot{\varepsilon}(p)+2v\cdot {\rm rot\,}
u\,.
 \end{align}
If $p,q \in {\cal Q}(\Omega)$ then
${\varepsilon}(p)={\varepsilon}(q)=0$, and these equalities imply
 \begin{align} \label{E3} & {\varepsilon}(pq)\,=({\rm div\,}
u)v-({\rm div\,} v)u - 2 \nabla_vu\,,\\
\label{E4} & (\alpha v + \beta u + u\wedge v)\,=\, \alpha \,{\rm
div\,} v+\beta\,{\rm div\,} u +2v\cdot {\rm rot\,} u\,.
 \end{align}
For $p,q \in {\dot{\cal Q}(\Omega)}$, we get
 \begin{align} \label{E5} & {\varepsilon}(pq)\,=\,- 2 \nabla_vu\,,\quad
{\rm div\,}(\alpha v + \beta u + u\wedge v)\,=\,2v\cdot {\rm
rot\,} u\,.
 \end{align}
\bigskip

\noindent{\bf Algebras ${\cal A}_e({\Omega})$.}\,\,\,Assume that
${\cal A} \subset{\cal Q}(\Omega)$ is an algebra and
$p=\{{\varphi},h\}\in {\cal A},\,\,h \not=0$. Such a field $p$ has
to possess the following properties.
\smallskip

\noindent{$\bullet$}\,\,\,The relation $\nabla{\varphi}={\rm
rot\,} h$ leads to ${\rm div\,} \nabla
{\varphi}=\Delta{\varphi}=0$, so that ${\varphi}$ is harmonic in
${\Omega}\setminus\Gamma$ and, hence, ${\varphi}\not=0$ almost
everywhere.
\smallskip

\noindent{$\bullet$}\,\,\,Since $p^2 \in{\cal A}$, one has
${\varepsilon}(p^2)=0$, and (\ref{E3}) for $q=p$ implies
$\nabla_hh=0$. Writing $h=\psi e$ with a smooth $\psi$ and
$|e|=1$, we have
 $$
0=\nabla_hh=\psi\left[(\nabla_e\psi)e+\psi\nabla_e e\right]\,,
 $$
($\nabla_e\psi:=e\cdot\nabla\psi$), whereas $e\cdot\nabla_e e=0$
holds. This implies
 \begin{equation}\label{psi e=0, nabla e e=0}
\nabla_e\psi=0\,,\quad\nabla_e e\,=\,0
 \end{equation}
and means that the lines of the vector field $h$ are the geodesics
and $|h|=|\psi|=\rm const$ along each line.
\smallskip

\noindent{$\bullet$}\,\,\,Since $p^2=\{{\varphi}^2-\psi^2,
2{\varphi}\psi e\}\in {\cal Q}(\Omega)$, the same arguments, which
have led to the first equality in (\ref{psi e=0, nabla e e=0}),
imply $\nabla_e[2{\varphi}\psi]=0$ and lead to
$\nabla_e{\varphi}=0$. Thus, we have
 \begin{equation}\label{psi e=phi e=0}
\nabla_e\psi\,=\,\nabla_e{\varphi}\,=\,0\,.
 \end{equation}
\smallskip

\noindent{$\bullet$}\,\,\,The scalar component of the quaternion
field $p^2\in {\cal Q}(\Omega)$ must be harmonic. Hence, we have
 \begin{align*}
&
0=\Delta({\varphi}^2-\psi^2)=2\left[{\varphi}\Delta{\varphi}+|\nabla{\varphi}|^2-\psi\Delta\psi-|\nabla\psi|^2\right]=\\
&
=\,2\left[|\nabla{\varphi}|^2-\psi\Delta\psi-|\nabla\psi|^2\right],
\quad {\rm i.e.,}\quad|\nabla{\varphi}|^2=\psi\Delta\psi+|\nabla
\psi|^2\,.
 \end{align*}
By the latter equality, with regard to ${\varphi}\not=\rm const$,
we conclude that $\psi\not=0$ almost everywhere. Indeed, assuming
the opposite, there is a set $A \subset \Omega$ such that ${\rm
mes}A>0$ and $\psi|_A=0$. Let $A^*\subset A$ be the set of density
points of $A$, so that ${\rm mes}A={\rm mes}A^*>0$
\footnote{Recall that $x \in A$ is a {\it density point} if
$\lim\limits_{r\to 0}\frac{{\rm mes}A\cap B_r[x]}{{\rm
mes}B_r[x]}=1$, where $B_r[x]$ is a ball of radius $r$ and center
$x$. If $A$ is measurable then almost every $x \in A$ is a density
point: see, e.g., \cite{Stein}.}. If $a \in A$ and $\nabla
\psi(a)\not=0$ then the set $A$ is a smooth surface near $a$ and,
hence, $a$ is not a density point of $A$. Hence, $\nabla
\psi|_{A^*}=0$, i.e., $\psi$ and $\nabla \psi$ vanish on $A^*$
simultaneously. Therefore,
$|\nabla{\varphi}|^2=\psi\Delta\psi+|\nabla \psi|^2=0$ on $A^*$.
In the mean time, $\nabla{\varphi}|_{A^*}=0$ for a harmonic
$\varphi$ yields $\varphi={\rm const}$ by general results of
elliptic theory (see, e.g., \cite{MV}). Thus, we arrive at a
contradiction. So, $\psi\not=0$ almost everywhere.

\smallskip

\noindent{$\bullet$}\,\,\,By the use of the third equality in
(\ref{E1}), we have
 \begin{equation}\label{*}
\nabla {\varphi}\,=\,{\rm rot\,} \psi e \,=\,\nabla\psi\wedge
e+\psi\,{\rm rot\,} e\,.
 \end{equation}
Multiplying by $e$, with regard to $e\perp [\nabla\psi\wedge e]$
and (\ref{psi e=phi e=0}), we arrive at the relation $e\cdot{\rm
rot\,} e=0$, which is a vectorial form of the Frobenius
integrability condition. Hence, we conclude that
 \begin{equation}\label{e=nabla tau}
e\,=\,\nabla \tau \qquad {\rm locally\,\,in}\,\,\,\Omega\,.
 \end{equation}
Therefore, {\it locally}, the level surfaces $S^c=\{x\in
{\Omega}\,|\,\,\tau(x)=c\},\,\,-\delta<c<\delta$ are geodesically
parallel, whereas $\tau(x)=\pm\,{\rm dist\,}(x,S_0)$.
\smallskip

\noindent{$\bullet$}\,\,\,By (\ref{e=nabla tau}), relation
(\ref{*}) takes the form $\nabla
{\varphi}=\nabla\psi\wedge\nabla\tau$, which is equivalent to
$\nabla \psi=\nabla\tau \wedge \nabla {\varphi}$. Applying ${\rm
div\,}$ to the latter, with regard to fourth equality in
(\ref{F1}) we get $\Delta \psi=0$. So, we have
 \begin{equation}\label{**}
\Delta {\varphi}\,=\,\Delta \psi\,=\,0\quad{\rm and}\quad \nabla
\psi\,=\,\nabla\tau \wedge \nabla {\varphi} \,\,\, {\rm
locally\,\,in}\,\,\,\Omega\,.
 \end{equation}
Thus, any $p \in {\cal A} \subset{\cal Q}(\Omega)$ is represented
in the form
 \begin{equation}\label{Fin Repres}
p\,=\,\left\{{\varphi}, \,\psi\nabla\tau\right\}\quad {\rm
locally\,\,in}\,\,\,\Omega
 \end{equation}
with $|\nabla\tau|=1$ and ${\varphi}, \psi$ satisfying (\ref{**}).

As is easy to verify, the converse is also true in the following
sense. If ${\cal Q}(\Omega)$ contains a field $p=\{{\varphi},h\}$
of the form (\ref{Fin Repres}) then there is a commutative algebra
${\cal A}_e({\Omega})\subset {\cal Q}(\Omega)$, which consists of
the elements $q=\{\lambda,\mu\nabla\tau\}$ and
 \begin{equation*}
pq\,=\,qp\,=\,\left\{{\varphi}\lambda-\psi\mu,
\,({\varphi}\mu+\psi\lambda)\nabla\tau\right\}
 \end{equation*}
holds.

Algebra ${\cal A}_e({\Omega})$ is evidently related with ${\cal
A}_e$ (\ref{A-h in C}), what motivates similarity in notation.
Moreover, it is closely related with the analytic function
algebras (\ref{CR 2d}). Indeed, let $p$ be of the form (\ref{Fin
Repres}). Define a complex-valued function $w={\varphi}+\psi i$;
let $w^c=w|_{S^c}={\varphi}^c+\psi^c i$ be its traces on the level
sets of $\tau$. Then, by virtue of the second equality in
(\ref{**}), ${\varphi}^c$ and $\psi^c$ turn out to be conjugated
by Cauchy-Riemann, in the same sense as in (\ref{CR 2d}).
Therefore, $w^c$ belongs to ${\cal A}(S^c)$, whereas ${\cal
A}_e({\Omega})$ is stratified as ${\cal
A}_e({\Omega})=\cup_{c}{\cal A}(S^c)$.
\smallskip

\noindent{$\bullet$}\,\,\,Assume in addition that $p\in{\cal
A}_e({\Omega})\cap{\dot{\cal Q}(\Omega)}$, so that ${\rm div\,}
\Im p={\rm div\,}\psi \nabla\tau=0$. In such a case, one has
 \begin{equation*}
0= {\rm div\,}\psi
\nabla\tau\overset{(\ref{F1})}=\nabla\psi\cdot\nabla\tau+\psi\Delta\tau\overset{(\ref{**})}=\psi\Delta
\tau
 \end{equation*}
and, hence, $\Delta\tau=0$ holds. Thus, if $p$ is a pure harmonic
field then the corresponding distant function $\tau$ is harmonic.
The converse is also true.
\smallskip

We say ${\cal A}_e({\Omega})$, as well as the corresponding
functions satisfying (\ref{**}) and associated with geodesic
vector fields $e$, to be {\it axial} algebras and  {\it axial}
harmonic functions ($e$ is an axis).
\smallskip

\noindent{$\bullet$}\,\,\,Elements of axial algebras obey the
maximum module principle: for their elements $p=\{{\varphi}, h\}$,
the relation
 \begin{equation}\label{max principle}
\max_{\Omega}|p|\,=\,\max_\Gamma|p|
 \end{equation}
is valid. Indeed, by $p\in{\cal A}_e({\Omega})$ one has
$\Delta{\varphi}=0,\,{\rm rot\,} {\rm rot\,} h=0$ and $\nabla_h
h=0$, which implies
 \begin{align*}
& \Delta|p|^2={\rm div\,}\nabla({\varphi}^2+h\cdot
h)\overset{(\ref{F1})}=2\left[{\varphi}\Delta{\varphi}+|\nabla{\varphi}|^2+{\rm
div\,}(\nabla_h
h+h\wedge{\rm rot\,} h)\right]=\\
& =2\left[|\nabla{\varphi}|^2+{\rm div\,}\nabla_h h+|{\rm rot\,}
h|^2 -h\wedge{\rm rot\,}{\rm rot\,}
h\right]=2\left[|\nabla{\varphi}|^2+|{\rm rot\,}
h|^2\right]\,>\,0\,.
 \end{align*}
Hence, $|p|^2$ is a subharmonic function and, as such, attains its
maximum at the boundary. Therefore, the same is valid for $|p|$.
\bigskip

\noindent{\bf Admissible metrics.}\,\,\,The question arises of
which $\Omega$ the axial algebras do exists. The exhausting answer
is not known, and the following is some considerations on this
point.
\smallskip

\noindent{$\bullet$}\,\,\,Here we provide an example of an algebra
${\cal A}_e({\Omega})$. Our construction is of local character.

Take a smooth surface $S \subset{\Omega}$. Let $\sigma^1,\sigma^2$
be the local coordinates on $S$ and $\tau:= \pm\,{\rm
dist\,}(\cdot, S)$. Near $S$ the semi-geodesic coordinates
$\sigma^1,\sigma^2,\tau$ are regular. Assume that the metric $g$
is of the form
 \begin{equation}\label{metric}
ds^2\,=\,d\tau^2\,+\,\rho(\tau)g_{ik}(\sigma^1,\sigma^2)\sigma^i\sigma^k
 \end{equation}
with a smooth positive conformal factor $\rho$. Choose two
functions ${\varphi}^0, \psi^0$ on $S$ related via the
CR-conditions $d\psi^0=\star d{\varphi}^0$ (w.r.t. the induced
metric $g|_S$). Extend them to a neighborhood in ${\Omega}$ by
${\varphi}(\sigma^1,\sigma^2,\tau)={\varphi}^0(\sigma^1,\sigma^2),\psi(\sigma^1,\sigma^2,\tau)\\=\psi^0(\sigma^1,\sigma^2)$.
As is easy to recognize, ${\varphi}$ and $\psi$ satisfy
(\ref{**}), whereas the field $p=\{{\varphi}, \psi \nabla\tau\}$
is harmonic and inscribed in the corresponding algebra ${\cal
A}_e({\Omega})$.

It is not improbable that this example exhausts all possible
cases. If so, in order for algebras ${\cal A}_e({\Omega})$ to
exist, the metric in ${\Omega}$ must be of the structure
(\ref{metric}) along the proper directions $e$'s. In particular,
such algebras do exist in the spaces of constant curvature
\footnote{S.V.Ivanov, private communication}.
\smallskip

\noindent{$\bullet$}\,\,\,As to a pure harmonic $p$, the condition
$\Delta \tau=0$ turns out to be very restrictive. For instance, it
cannot be realized in the 3d sphere $S^3$ even locally
\footnote{S.V.Ivanov, private communication}.

\section{${\mathbb H}$-spectrum}
{\bf Axial algebras in ${\mathbb R}^3$.}\,\,\,Let $\Omega
\subset{\mathbb R}^3$ be an (open) bounded domain with the smooth
boundary $\Gamma$. In this case, there is a rich reserve of
algebras ${\cal A}_e({\Omega})$.
\smallskip

\noindent{$\bullet$}\,\,\,Fix an $O\in {\mathbb R}^3$ and the
polar coordinate system $\phi, \theta, r$ with the pole $O$.
Recall that, in the polar coordinates, the ${\mathbb R}^3$-metric
takes the form (\ref{metric}).

For a point $x$, by $\vec r(x)$ we denote its radius-vector
applied at $x$, so that $|\vec r|=r$. Thus, we have a
spherically-symmetric geodesic field $e=r^{-1}\vec r$ with ${\rm
div\,} e(x)=2 r^{-1}$.

Denote $S^c_O=\{x \in {\mathbb R}^3\,|\,\,r(x)={\rm
dist\,}(x,O)=c\}$. For a point $x=x(\phi, \theta, r)\not= O$,
denote by $\pi(x)\in S^1_O$ its geodesic projection on the unit
sphere, i.e., the point with the coordinates $(\phi, \theta, 1)$.

Let ${\Omega}$ and $O$ be such that the chosen polar system is
regular in a neighborhood of $\bar{\Omega}$. In this case, the
domain is regularly stratified: $\bar{\Omega}=\cup_c
[\bar{\Omega}\cap S^c_O]$.

Let ${\varphi}^1, \psi^1$ be two functions on $S^1_O$ (of
variables $\phi, \theta$) continuous in $\pi(\bar{\Omega})$ and
provided $d\psi^1=\star\,d{\varphi}^1$ in $\pi({\Omega})$. In
$\bar{\Omega}$, define the functions
${\varphi}={\varphi}^1(\pi(x)), \psi=\psi^1(\pi(x))$. Then, just
by construction, the quaternion field $p=\{{\varphi}, \psi e\}$
turns out to be harmonic and belongs to ${\cal Q}(\Omega)$. The
fields of this form constitute the axial algebra ${\cal
A}_e({\Omega})$. In the mean time, the constructed $p$'s are not
pure harmonic, i.e., ${\cal A}_e({\Omega})\not\subset{\dot{\cal
Q}(\Omega)}$ since ${\rm div\,} \psi e=\nabla\psi\cdot e+\psi{\rm
div\,} e\overset{(\ref{psi e=0, nabla e e=0})}=\psi 2
r^{-1}\not=0$.

For a fixed $\Omega$, varying properly the position of the pole
$O$, we "prospect" the domain by the radial fields $e$ and get a
rich family of the algebras ${\cal A}_e({\Omega})\subset {\cal
Q}(\Omega)$.
\smallskip

\noindent{$\bullet$}\,\,\,Such a family becomes even richer if one
changes spheres by planes. Fix an $\omega,\,\,|\omega|=1$, and
denote $\Pi^c_\omega=\{x \in {\mathbb R}^3\,|\,\,x\cdot
\omega=c\}$. Then the domain is stratified as
$\bar{\Omega}=\cup_c[\bar{\Omega} \cap \Pi^c_\omega]$, whereas
$\pi(\bar{\Omega})=\{x-(x\cdot\omega)\omega\,|\,\,x\in \bar
{\Omega}\}$ is its projection onto $\Pi^0_\omega$.

Choose two functions ${\varphi}^0, \psi^0$ on $\Pi^0_\omega$
continuous in $\pi(\bar{\Omega})$ and such that
$d\psi^0=\star\,d{\varphi}^0$ in $\pi(\Omega)$. In $\bar{\Omega}$,
define the functions ${\varphi}={\varphi}^0(\pi(x)),
\psi=\psi^0(\pi(x))$. The quaternion field $p=\{{\varphi}, \psi
e\}$ is harmonic and belongs to ${\cal Q}(\Omega)$. Such fields
constitute the axial algebra ${\cal A}_\omega({\Omega})$ (here
$\omega$ is understood as a constant vector field $e\equiv \omega$
in ${\mathbb R}^3$). Moreover, by virtue of ${\rm div\,}
\omega=0$, its elements turn out to be pure harmonic, so that
${\cal A}_\omega({\Omega})\subset{\dot{\cal Q}(\Omega)}$ holds.

Thus, there is the family of pure harmonic axial algebras indexed
by the unit vectors $\omega$.
\smallskip

\noindent{$\bullet$}\,\,\,By calculations quite analogous to the
ones which have led to (\ref{max principle}), one can show that
all elements of the space ${\dot{\cal Q}(\Omega)}$ obey the
maximum module principle.
\bigskip

\noindent{\bf AH-structure on ${\dot{\cal Q}(\Omega)}$.}\,\,\,A
specific feature of the case ${\Omega}\subset{\mathbb R}^3$ is
that the quaternion fields are canonically identified with the
${\mathbb H}$-fields (${\mathbb H}$-valued functions) by the
correspondence $\{\alpha, u\}\equiv \alpha+(u\cdot e_1){\bf
i}+(u\cdot e_2){\bf j}+(u\cdot e_3){\bf k}$, where
$e_1,\,e_2,\,e_3$ is a fixed orthonormal basis in ${\mathbb R}^3$.
As a consequence, the space ${\dot{\cal Q}(\Omega)}$ is provided
with some additional algebraic structure. We describe it, keeping
the notions and terminology of the paper \cite{Joyce}.
\smallskip

\noindent $\bullet$\,\,\,Begin with a portion of abstract
definitions.

Let a real linear space ${\cal U}$ be a (left) ${\mathbb
H}$-module with the action $u \mapsto au$ for $u \in {\cal U},\,\,
a \in {\mathbb H}$.

By ${\cal U}^\times$ we denote the ${\mathbb H}$-dual space, i.e.,
the space of linear maps $f: {\cal U}\to {\mathbb H}$, which
satisfy $f(ap)=af(p)$, $a \in {\mathbb H}$. Let us call such $f$'s
{\it ${\mathbb H}$-functionals}. Note that ${\cal U}^\times$ is a
left ${\mathbb H}$-module with the action $(bf)(p)=f(p)\bar
b,\,\,b \in {\mathbb H}$. If ${\cal U}$ is a normed space then
${\cal U}^\times$ is also endowed with the norm
$\|f\|_\times=\sup_{\|u\|=1}|f(u)|$.

Let ${\cal U}^\prime \subset {\cal U}$ be a subspace. Note that we
do not require ${\cal U}^\prime$ to be invariant w.r.t. ${\mathbb
H}$-action. Define ${\cal U}^\dagger =\{f \in {\cal
U}^\times\,|\,\,f(u)\in {\mathbb I}\,\,\,{\rm
for\,\,all}\,\,\,u\in{\cal U}^\prime\}$.

We say the pair $\{{\cal U},{\cal U}^\prime\}$ to be an {\it
AH-module} (augmented ${\mathbb H}$-module) if $f(u)=0$ for all $f
\in {\cal U}^\dagger$ implies $u=0$. This means that the subspace
${\cal U}^\dagger$ possesses a {\it totality} property: it
distinguishes elements of ${\cal U}$.
\smallskip

\noindent $\bullet$\,\,\,Now, let us show that in ${\mathbb R}^3$
the pure harmonic quaternion fields constitute an AH-module.

Define the action of ${\mathbb H}$ on ${\dot{\cal Q}(\Omega)}$.
Fix an $a\in {\mathbb H}$ and denote by $\tilde a(\cdot)=a$ the
constant quaternion field in ${\Omega}$. Now, for
$p=\{{\varphi},h\} \in {\dot{\cal Q}(\Omega)}$ we put $ap=\tilde
a(\cdot)p(\cdot)$ (point-wise). By (\ref{E5}), one has
\begin{align*} \label{E5} & {\varepsilon}(ap)=- 2 \nabla_h\Im\tilde a=0\,,\quad
{\rm div\,}\Im (ap)\,=\,2h\cdot {\rm rot\,} \Im\tilde a=0\,,
 \end{align*}
just because $\tilde a$ is constant. Hence, $ap \in {\dot{\cal
Q}(\Omega)}$, so that the ${\mathbb H}$-action is well defined.
Thus, ${\dot{\cal Q}(\Omega)}$ is an ${\mathbb H}$-module.

Let $[{\dot {\cal Q}}(\Omega)]^\times$ be the ${\mathbb H}$-dual
space.

Take $[{\dot {\cal Q}}(\Omega)]^\prime=\{p \in {\dot{\cal
Q}(\Omega)}\,|\,\,\Re p=0\}={\dot{\cal Q}(\Omega)}\cap{\cal
I}(\Omega)$. This subspace consists of the fields $p=\{0,h\}$ such
that ${\rm div\,} h=0$ and ${\rm rot\,} h=0$. Recall that $[{\dot
{\cal Q}}(\Omega)]^\dagger=\{f \in[{\dot {\cal
Q}}(\Omega)]^\times\,|\,\,f(u)\in {\mathbb I}\,\,{\rm for\,\,
all}\,\,\,u\in[{\dot {\cal Q}}(\Omega)]^\prime\}$. This subspace
contains the {\it quaternion Dirac measures} $\theta_m$, which are
associated with points $m\in\bar{\Omega}$ and act by
$\theta_m(p)=p(m)\in {\mathbb H}$. These measures distinguish
elements of ${\dot{\cal Q}(\Omega)}$: if $\theta_m(p)=0$ for all
$m$ then $p=0$. Hence, moreover, the wider set $[{\dot {\cal
Q}}(\Omega)]^\dagger$ does distinguish elements, i.e., possesses
the totality property.

Summarizing, we conclude that $\{{\dot{\cal Q}(\Omega)},[{\dot
{\cal Q}}(\Omega)]^\prime\}$ is an AH-module.
\smallskip

\noindent $\bullet$\,\,\,Denote
$\Theta({\Omega})=\{\theta_m\,|\,\,m \in \bar{\Omega}\}\subset
[{\dot {\cal Q}}(\Omega)]^\times$. For any $\theta_m$ and $p \in
{\dot{\cal Q}(\Omega)}$, one has
 $$
|\theta_m(p)|=|p(m)| \leqslant \sup_{\Omega}
|p(\cdot)|=\|p\|_{\dot{\cal Q}(\Omega)}
 $$
that implies $\|\theta_m\|_\times\leqslant 1$. In the mean time,
the $\sup$ is attained on $p=\{1,0\}$. Hence, $\|\theta_m\|_\times
= 1$, i.e., $\Theta({\Omega})$ is embedded to the unit sphere of
the dual space $[{\dot {\cal Q}}(\Omega)]^\times$. This sphere is
compact w.r.t. the $*$-topology determined by the point-wise
convergence of ${\mathbb H}$-functionals on elements of
${\dot{\cal Q}(\Omega)}$. As one can show, in this topology
$\Theta({\Omega})$ is a compact set. Moreover, the bijection
$\Theta({\Omega})\ni \theta_m \leftrightarrow m \in \bar{\Omega}$
is a homeomorphism of topological spaces.

Choose a unit vector $\omega$; let $y, z \in {\cal
A}_\omega({\Omega})$. Recall that ${\cal A}_\omega({\Omega})$ is a
commutative algebra. For any $\theta_m$ one has
 $$
\theta_m(yz)=(yz)(m)=y(m)z(m)=\theta_m(y)\theta_m(z)\,,
 $$
i.e., the ${\mathbb H}$-functional $\theta_m$ is {\it
multiplicative} on ${\cal A}_\omega({\Omega})$.
\smallskip

\noindent $\bullet$\,\,\, The above mentioned properties of the
Dirac measures characterize the set $\Theta({\Omega})$. Namely,
let us define ${{\Omega}}^{\mathbb H} \subset [{\dot {\cal
Q}}(\Omega)]^\times$ as a set of ${\mathbb H}$-functionals of the
norm $1$, which act multiplicatively on each algebra ${\cal
A}_\omega({\Omega})$. Then, modifying properly the arguments of
\cite{Joyce} (section 3.4), one can show that ${\Omega}^{\mathbb
H}=\Theta({\Omega})$.

We say ${{\Omega}}^{\mathbb H}$ to be an {\it ${\mathbb
H}$-spectrum} of the domain ${\Omega}\subset {\mathbb R}^3$. As it
follows from the aforesaid, the ${\mathbb H}$-spectrum is
homeomorphic to the domain.
\smallskip

\noindent $\bullet$\,\,\,By Gelfand, the null-subspace ${\rm
Ker\,}[\theta_m|_{{\cal A}_\omega({\Omega})}]$ corresponds to the
maximal ideal in ${\cal A}_\omega({\Omega})$, which consists of
(axial) analytic functions $w={\varphi}+\psi i$ vanishing on the
straight line, which passes through $m \in \bar {\Omega}$ in
parallel to $\omega$. This line intersects the projection
$\pi(\bar{\Omega})$ of the domain $\bar {\Omega}$ onto the
orthogonal plane $\Pi^0_\omega$ at the point $\pi(m)$. As a
result, also by Gelfand, each algebra ${\cal A}_\omega({\Omega})$
determines this projection up to homeomorphism. Moreover, it
determines $\pi(\bar{\Omega})$ (as a 2d Riemannian manifold) up to
conformal equivalence \cite{BCald}.
\bigskip

\noindent{\bf Comments and conjectures.}\,\,\,$\bullet$\,\,\,In
the generic case of 3d ${\Omega}$, the vector parts $u$ of the
fields $p=\{{\varphi},u\}\in {\cal Q}(\Omega)$ take values in
tangent spaces $T\Omega_m$ but not in ${\mathbb R}^3$, and no
canonical way is seen to identify fields $p$ with ${\mathbb
H}$-values functions (as in the case of ${\Omega}\subset {\mathbb
R}^3$). Surely, one can turn ${\mathbb R}\times T\Omega_m$ into
${\mathbb H}$ by choosing a local frame but the analog of the
constant fields $\tilde a$, which determine the action of $\mathbb
H$ on ${\cal Q}(\Omega)$, does not appear. As a consequence, the
AH-structure disappears, what reduces the options of the algebraic
approach to ITP.

However, one can define the ${\mathbb H}$-spectrum as follows.
Beginning with the space ${\cal Q}(\Omega)$, we introduce the dual
space $[{\cal Q}(\Omega)]^\times$ of $\mathbb R$-linear operators
from ${\cal Q}(\Omega)$ to ${\mathbb H}$. Then, we {\it define}
$\Omega^{\mathbb H}\subset [{\cal Q}(\Omega)]^\times$ as the unit
norm elements, which act multiplicatively on the axial algebras
${\cal A}_e(\Omega)\subset{\cal Q}(\Omega)$ (if the latter do
exist; otherwise, we put $\Omega^{\mathbb H}=\emptyset$). However,
the question arises whether such a definition is rich in content.
In particular, can one hope for the homeomorphism $\Omega^{\mathbb
H}\cong\Omega$?

Perhaps, a relevant general definition of ${\Omega}^{\mathbb H}$
may be extracted from the D.Quillen paper \cite{Qui}, which
interprets D.Joyce's constructions in algebraic geometry terms.
Unfortunately, the author of the given paper is not quite educated
in Algebra to understand what is written there.
\smallskip

\noindent $\bullet$\,\,\,There is a case, in which ${\cal
Q}(\Omega)$ contains at least one algebra ${\cal A}_e({\Omega})$,
whereas the ${\mathbb H}$-spectrum is well defined and available
for reconstruction of $\Omega$. Let the manifold be {\it
cylindric}, i.e., ${\Omega}=M\times[0,1]$, whereas the metric $g$
is of the form (\ref{metric}) with $\rho=\rm const$. Then there is
the algebra ${\cal A}_e({\Omega})$ with the axial field $e$, which
has the lines $\{m\}\times [0,1],\,\,m \in M$. In such a
situation, the "cross-section" $M$ can be recovered as the Gelfand
spectrum $M^{\mathbb C}$ of algebra ${\cal A}_e({\Omega})$ (even
if $g_M$ is not a simple metric). If ${\cal A}_e({\Omega})$ is a
unique commutative algebra in ${\cal Q}(\Omega)$ then
$\Omega^{\mathbb H}\cong M^{\mathbb C}$ holds.

In this connection, note that just a combination of the results of
\cite{BCald,BCUBO} and the given paper leads to the following
assertion on the uniqueness in ITP.

  \begin{Proposition}
Let a 3d manifold $(\Omega,g^\dagger)$ be such that
  \begin{enumerate}
 \item $\Omega=M\times I$, where $M$ is a simply connected compact 2d manifold
 with boundary, $I=[0,1]$;
 \item the metric $g^\dagger$ is of the form (\ref{metric}) with $\rho\equiv1,\,\,0\leqslant\tau\leqslant 1$
($\sigma^1,\sigma^2$ are the local coordinates on $M$).
 \end{enumerate}
Let $g$ be a metric on $\Omega$, $\Lambda_g$ and
$\Lambda_{g^\dagger}$ the DN-operators, corresponding to the
metrics $g$ and $g^\dagger$. In such a case, if $\Lambda_g =
\Lambda_{g^\dagger}$, then $g=\Phi_*g^\dagger$, where $\Phi:
\Omega \to \Omega$ is a diffeomorphism, which acts identically on
$\partial \Omega$.
  \end{Proposition}
\noindent {\it Proof (scetch).}

\noindent$\square$\,\,Since $M$ is simply connected, $\Omega$ is
also simply connected (as a 3d manifold). Therefore, $\Lambda_g$
determines the operator $\vec\Lambda_g$ (a "magneto-static"
DN-map: see \cite{BCUBO}) .

Given the pair $\Lambda_g, \vec\Lambda_g$ one can determine the
traces of the fields $p \in {\dot{\cal Q}}(\Omega)$ at $\partial
\Omega$ \cite{BCUBO,BShar}. Checking whether the trace of $p^2$
belongs to ${\dot{\cal Q}}(\Omega)|_{\partial \Omega}$, one can
select the "algebraic" $p$'s and determine the traces of all the
algebras ${\cal A}_e(\Omega)$.

The traces of elements of the algebra ${\cal A}_\omega(\Omega)$
corresponding to the field of geodesics $\left\{\{m\}\times
I\,|\,\,m \in M\right\}$ are of specific form. As such, they can
be selected and provide the trace algebra ${\cal
A}_\omega(\Omega)|_{\partial \Omega}$.

Owing to the maximum module principle, the latter algebra
determines the "invisible" algebra ${\cal A}_\omega(\Omega)$ up to
isometric isomorphism. Hence, we can find the Gelfand spectrum of
${\cal A}_\omega(\Omega)|_{\partial \Omega}$ and, thus, determine
$(M,\lambda g_M)$, where $\lambda=\lambda(m)$ is an (unknown)
conformal factor \cite{BCald}.

The symbol of $\Lambda_g$ determines $g|_{\partial \Omega}$
(G.Uhlmann et al). In particular, it determines
$g|_{M\times\{0\}}$ and, hence, fixes the factor $\lambda$. Thus,
the metric $g^\dagger$ is recovered.\,\,\,$\blacksquare$
\smallskip

So, at least in the dimension 3, the assumption on the injectivity
of the ray transform accepted in \cite{FKLS} is unnecessary. Note
that if $\vec\Lambda$ is given in addition to $\Lambda$, then the
assumption on $M$ to be simply connected can be also cancelled.

A curious point is that, at the current stage of understanding the
ITP, the class of admissible metrics is the same for the
geometrical optic approach and our algebraic approach. Perhaps,
there are some intimate relations between GO-solutions and
algebras ${\cal A}_e(\Omega)$.
\smallskip

\noindent $\bullet$\,\,\, It is not improbable that there are the
cases, when a {\it finite} number of the axial algebras do exist
and is available for reconstruction.  Also, perhaps, the
reconstruction by \cite{Joyce}, where the author deals with the
even-dimensional hypercomplex manifolds, may be adapted for a
class of the impedance tomography problems. We plan to touch upon
this subject in future papers.

\end{document}